\documentclass[aps,pra,reprint,superscriptaddress,floatfix]{revtex4-1}

\usepackage{graphicx}
\usepackage{amsmath}
\usepackage{xcolor}

\begin{document}

\title{Koopmans' condition in self-interaction corrected density functional theory}


\author{P. Kl\"upfel}
\affiliation{CNRS, Laboratoire Collisions Agr\'egats R\'eactivit\'e (IRSAMC), F-31062 Toulouse, France}
\affiliation{CNRS, Laboratoire de Physique Th\'eorique (IRSAMC), F-31062 Toulouse, France}
\affiliation{Icelandic Center of Computational Science, University of Iceland,  107 Reykjav{\'i}k, Iceland}

\author{P. M. Dinh}
\affiliation{CNRS, Laboratoire de Physique Th\'eorique (IRSAMC), F-31062 Toulouse, France}
\affiliation{Universit\'e de Toulouse, UPS, Laboratoire de Physique Th\'eorique (IRSAMC), F-31062 Toulouse, France}

\author{P.-G. Reinhard}
\affiliation{Institut f\"ur Theoretische Physik, Universit{\"a}t Erlangen, D-91058 Erlangen, Germany}
\author{E. Suraud}
\affiliation{CNRS, Laboratoire de Physique Th\'eorique (IRSAMC), F-31062 Toulouse, France}
\affiliation{Universit\'e de Toulouse, UPS, Laboratoire de Physique Th\'eorique (IRSAMC), F-31062 Toulouse, France}

\date{\today}

\begin{abstract}
We investigate from a practitioner's point of view the computation of
the ionization potential (IP) within density functional theory
(DFT). DFT with (semi-)local energy-density functionals is plagued by
a self-interaction error which hampers the computation of IP from the
single-particle energy of the highest occupied molecular orbital
(HOMO). The problem may be cured by a self interaction correction
(SIC) for which there exist various approximate treatments. We compare
the performance of the SIC proposed by Perdew and Zunger with the
very simple average-density SIC (ADSIC) for a large variety of atoms
and molecules up to larger systems as carbon rings and chains.
Both approaches to SIC provide a large improvement to the quality of the
IP if calculated from the HOMO level. The surprising result is that the 
simple ADSIC performs even better than the original Perdew-Zunger SIC 
(PZSIC) in the majority of the studied cases. 
\end{abstract}


\maketitle

\section{Introduction\label{sec:intro}}

Density Functional Theory (DFT)~\cite{Par89,Dre90,Koh99r} has become a
standard theoretical tool for the investigation of electronic
properties in many physical and chemical systems. It provides fairly
reliable results with moderate computational effort.  Practical
implementations of DFT employ simple and robust approximations for the
exchange and correlation functional. The simplest one is the Local Density
Approximation (LDA), which has been proven very useful in calculations 
of electronic structure and dynamics. Typical applications are reaching 
from first principle calculations on the electronic ground state and molecular 
geometries \cite{Par94}, over dynamic studies of near equilibrium situations 
(e.g. optical response, direct one-photon processes) to highly 
non-linear dynamical scenarios \cite{Rei03a,Mar06aB,Fen08a}. 
However, LDA is plagued by a self-interaction error \cite{Per81}: The Kohn-Sham 
(KS) mean field is computed from the total density which includes all 
occupied single-particle states, including the state on which the
LDA field actually acts. 
Locality of the energy functional leads to a wrong asymptotic KS field.
This is still a great hindrance in many applications, for instance a
possibly large underestimation of IP and the absence of Rydberg or excitonic series in the
static KS spectrum~\cite{Sho77,Sch78}, the polarizability in chain molecules
\cite{Gis99a,Kue04a} or the spectral and fundamental gap in solids \cite{Hyb86a,Nie00aR}. Another
challenging application is the description of molecules or clusters deposited on surfaces
\cite{Bae07a,Din09aR}. In dynamic scenarios, the self-interaction error also dramatically
affects ionization dynamics, especially close to thresholds, e.g., in a time-dependent DFT
model of electron emission \cite{Poh00,Poh04b,Fen08a,Gio12a}.

In practice, the wrong asymptotics of the KS field stems from an incomplete
cancellation of the self-interaction error between the Hartree potential and the approximate
exchange and correlation field. 
Such a spurious
self-interaction is avoided by a complicated non-locality in exact KS-DFT
\cite{Per82,Per83,Sha83}. For the total energy, the requirement for non-locality can be
incorporated into gradients of density leading to the generalized gradient approximation
(GGA) \cite{Bec88, Per91, Per96}. This approximation indeed served to lift DFT to a quantitative level
in molecular physics and chemistry problems, but is insufficient to restore proper
asymptotics of the mean field. 

Although there are approaches to improve the asymptotic KS potential
\cite{Cas00a} those are often too demanding for practical
calculations, in particular in the time-domain. The aforementioned
examples show that there is still a need for robust and practical
approaches to improve on the asymptotic KS potential, such as self-interaction correction (SIC).
 
The original proposal for a SIC \cite{Per79,Per81} by Perdew and Zunger (PZ) has been
developed at various levels of refinements and proved to be useful over the years, in particular
for structure calculations in atomic, molecular, cluster and solid state physics, see e.g.
\cite{Ped84,Goe97,Polo,Vyd04,Hof12,Klu11,Klu12b,Sva96}. This original PZSIC scheme,
however, leads to an orbital-dependent mean field which causes several formal and
technical difficulties \cite{Goe97, Sva96, Mes08, Klu11, Hof12}. There are attempts to
circumvent the orbital dependence by treating SIC with optimized effective potentials
(OEP)\cite{Sha34}, for a review see \cite{Kue08}. However the resulting formalism is,
again, quite involved and usually treated approximately \cite{Kri92}. A very robust and
simple SIC is the average density SIC (ADSIC) which was
proposed already very early \cite{Fer34}, taken up in \cite{Leg02a}, and has been used
since in many applications to cluster structure and dynamics in all regimes.

ADSIC takes the total density divided by the electron number as a
reference for the single electron density in each state. Non-locality is incorporated
in the scheme by the global density integral providing the total particle 
number. However the
ADSIC functional, unlike the PZSIC one, is a functional of the total
density and thus, the subsequent mean field is not orbital dependent
anymore. Having the correct total charge, ADSIC provides the proper
asymptotics for the mean field. It is argued that the approximation by
one and the same (average) single-particle density for each state is only
applicable in simple metals where all electronic states cover the same
region of space, that is, in the case of a cluster, the whole cluster itself \cite{Leg02, Leg02a}. 
Later studies revealed that ADSIC is also an efficient correction
scheme for non-metallic systems with delocalized electrons, such as organic 
molecules \cite{Ila03, Ila05}. 

The aim of this paper is to investigate the performance of ADSIC in
direct comparison to PZSIC for a large variety of atoms and molecules in 
their ground state. The sample covers systems of different binding types, and
not only metallic ones. We will compare ADSIC with a mere DFT treatment 
using (semi-)local functionals and with PZSIC, also occasionally with
Hartree-Fock. The comparison focuses on the proper description of the
IP. We start from atoms as elementary building
blocks of any molecule, proceed to a large variety of molecules from
simple dimers to more complex organic structures, and finally discuss
carbon rings and chains with a systematic variation of sizes. Such a 
strategy allows us to cover various binding types but also various 
geometries and even dimensionality.

\section{Ionization potential \label{sec:ionenerg}}

\subsection{Definitions}

The key quantity of this survey is the ionization energy $I$, commonly
called ionization potential (IP). 
The IP of a $N$ electron system is given by the energy 
difference
\begin{equation}\label{eqn:IP_ediff}
I \equiv I_{\Delta} = E(N-1) - E(N) \ .
\end{equation}
The energies $E(p)$ correspond to ground-state configurations of a
$p$ particle system in a given external potential, typically the Coulomb potential created
by the nuclear charges. Both energies, $E(N)$ as well as $E(N\!-\!1)$, are fundamental
observables in ground-state DFT and so is its difference, the IP. DFT should thus allow
one to calculate the IP of electronic systems. A distinction has to be made here. The
definition of an IP is unique in atoms. In molecules, however, we distinguish vertical and
horizontal IP. The vertical one considers the energy difference from the removal of one
electron for frozen atomic positions. This typically corresponds to photon induced fast
emission processes. The horizontal IP is built from the difference of fully relaxed
molecular configurations. It accounts for the energy change on a long time scale on which
all molecular relaxation processes are finalized. We will consider throughout this paper
the vertical IP which accounts for fast electronic processes and which is closely related
to the properties of the electronic ground state of the mother system, in particular to
the highest occupied molecular orbital (HOMO).

In the exact electronic ground state, the asymptotic decrease of the
ground state density $n$ is related to the IP by
\begin{equation}
  n({\bf r}) 
  \stackrel{|{\bf r}| \to \infty}{\sim} 
  \exp\left[-2\sqrt{2I}|{\bf r}|\right] \ .
\end{equation}
In an exact KS-DFT, the asymptotic decay of the total density is
defined merely by the highest occupied KS orbital, i.e., the HOMO 
\cite{Par94}.
In combination with the proper asymptotic behavior of the KS potential 
($v_s({\bf r})\to 0$), the ionization energy can thus be related to the 
single-particle energy of the HOMO~:
\begin{equation}\label{eqn:IP_HOMO}
I \equiv I_{\varepsilon} = -\varepsilon_{\rm HOMO} \ .
\end{equation}
For an exact exchange-correlation functional, both definitions of the
ionization energy, i.e., Eqs. (\ref{eqn:IP_ediff}) and
(\ref{eqn:IP_HOMO}), coincide, i.e. they obey
$I_\Delta=-\varepsilon_{\rm HOMO}$. The identification of the negative
HOMO energy with the IP was referred to as Koopmans'
theorem \cite{Koo33} long before the fundamental concepts where
extended rigorously to the framework of DFT \cite{Hoh64,Koh65}. For
approximate energy functionals Koopmans' condition does not
necessarily hold \cite{Dre90} and the deviation from the ideal
behavior can be used to define the Non-Koopmans (NK) energy
\begin{equation}\label{eqn:NK}
\Delta E_{\rm NK} = I_{\varepsilon} - I_{\Delta} \ . 
\end{equation}
A value $\Delta E_{\rm NK}=0$ signals that Koopmans' theorem is
fulfilled. In such a situation, the properties of the HOMO level
are closely related to ionization and electron emission. 
We know that LDA produces rather large violations of Koopmans' theorem
and thus exhibits sizable  $\Delta E_{\rm NK}$. SIC should reduce that, and
the amount of reduction is one measure of the efficiency of the actual
SIC scheme. 

It is also interesting to compare the performance of a calculation
with respect to data. Thus we consider in addition to the NK energy
the bare error in the IP relative to experiments or other reference data. 
As we have two 
definitions of the IP, there are two bare errors in an approximate theory~:
\begin{equation}\label{eqn:errors}
\Delta I_{\varepsilon} = I_{\varepsilon} - I_{\rm ref}
\ , \qquad
\Delta I_{\Delta} = I_{\Delta} - I_{\rm ref}
\;.
\end{equation}
An experimental reference energy may be hampered by uncertainties, as
ionic relaxation throughout the ionization process can lead to
situations which lie between the idealized vertical IP (for very fast
ionization) and horizontal one (for very slow
ionization). Here only vertical IP in the ground-state geometry is 
considered. Reliable atomic coordinates for small molecules are given 
by the MP2 optimized structures as provided in the G2 dataset \cite{Cur97}.
G2 theoretical and experimental energies may differ by several tens 
to hundrets of meV \cite{Cur91}. This error can be considered negligible 
on the scale of the expected errors, stemming from the approximate nature of the 
used exchange-correlation functionals and the use of pseudo-potentials \cite{Goe98}. 
Experimental data for vertical IP therefore appear as a safe choice of reference \cite{cccbdb}.

In contrast to the errors in the IP (\ref{eqn:errors}), the NK energy does 
not require any reference data that may be hampered by experimental 
uncertainties. It therefore provides a rather rigorous 
criterion for the quality of energy functional approximations.

\subsection{Impact of a proper description of IP}

The two definitions (\ref{eqn:IP_ediff}) and (\ref{eqn:IP_HOMO}) for
the IP are equally justified in an exact calculation.  However, for
(semi-)local functionals (as in LDA and GGA), it is usually found that only the
energy difference (\ref{eqn:IP_ediff}) can be used to extract a good estimate for the IP.
The estimate (\ref{eqn:IP_HOMO}) from the single-particle spectrum
requires the proper $1/r$-asymptotics of the exchange-correlation
potential (for neutral systems). This is not provided in calculations
based on typical semi-local functionals.

While energy differences often allow reliable estimates already with
semi-local functionals, they require two calculations, which is more
involved than a straightforward extraction from the HOMO level. This
alone would not be {\it a priori} a major hindrance. But there are many situations
where the extraction via energy difference is not an option~: In
periodic calculations (e.g., on surfaces), a rigorous calculation of
the IP (called work function in this case) from an energy difference is hard
to achieve because one cannot easily model a single excess charge in a
periodic setup. The same situation applies for calculation of band-gaps in
solids. In dynamical situations, as described by time-dependent DFT, 
an accurate modeling of the ionization
process requires an accurate static single-particle spectrum. As the
propagation of the single-particle states is driven by the
time-dependent KS Hamiltonian, the energy differences in its spectrum
and a proper position of the IP is more essential than the total
energy. Ionization properties are also mostly defined 
by the HOMO level which thus has to be correctly described.

\begin{figure}[htbp]
\includegraphics[width=0.85\columnwidth]{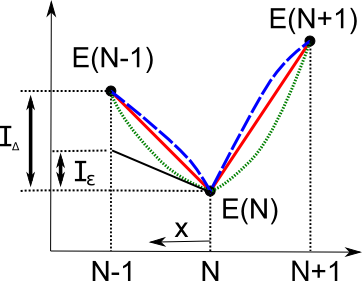}
\caption{\label{fig:cartoon_IP}(Color online) Illustration of the ground state energy
  $E$ as a function of fractional occupation number $\nu$. The IP
  from the HOMO level, $I_\varepsilon$, corresponds to the left-handed 
  derivative (slope = black line) of the energy at $\nu=N$. The IP from energy differences $I_\Delta$,
  is associated with the exactly linear behavior (red/solid line).  The
  LDA result provides a smooth curve (green/dotted line). Hartree-Fock 
  (blue/dashed line) also has a discontinuous derivative at $\nu=N$ as 
  the exact trend, but tends to overestimate the kink.}
\end{figure}

A way to illustrate the self-interaction error is to consider the
energy $E(\nu)$ as a function of a fractional particle number $\nu$. The
ionization process goes along $\nu=N\longrightarrow N\!-\!1$.  An exact
functional produces a linear behavior \cite{Per82}, as~:
\begin{equation}
E(\nu) = (1\!-\!x) E(N) + xE(N\!-\!1) \ ,\\
\ 
x=N-\nu\in[0,1]
\,.
\label{eq:exactElin}
\end{equation}
A similar linear behavior is also observed for ionization from an
anion to the neutral system ($N+1\longrightarrow N$).
This exact $E(\nu)$ is shown as (red) solid line in Figure \ref{fig:cartoon_IP}.
The remarkable feature is a discontinuous derivative, that is a kink, at $\nu=N$.
Semi-local functionals deal with smooth functions (no kinks, no discontinuities) 
and produce smooth trends as shown in the (green) dotted line.
The definition (\ref{eqn:IP_HOMO}) of the IP through the HOMO energy
coincides with the slope of the total energy for fractional particle
numbers at $\nu\nearrow N$:
\begin{equation}
I_{\varepsilon} = -\left.\frac{dE}{d\nu}\right|_{\nu\nearrow N}
\end{equation}
as indicated in the figure. The linear trend of the exact energy
 naturally guarantees $I_{\varepsilon}=I_\Delta$, while the convex curve
from LDA necessarily implies $I_{\varepsilon}<I_\Delta$.

The (blue) dashed line in Figure \ref{fig:cartoon_IP} finally shows 
results from exact exchange in Hartree-Fock (HF). Full HF is free from
self-interaction error. Thus it qualitatively yields the correct
result, namely the kink at $\nu=N$. It however differs from the linear
trend in between the integer particle numbers. This leads to an
overestimation of the IP from the HOMO, $I_{\varepsilon}>I_\Delta$.
We will address this question in the last example of carbon chains, see
 Figure \ref{fig:formation_carbon}.

\section{Various schemes for a SIC}

As SIC is rather an {\it ad-hoc} measure to cure the self-interaction
problem, various recipes and approximations are used, depending on the
field of application.  In this section, we briefly summarize the 
PZSIC and ADSIC which we will use later on in the extensive comparison 
of results. 

\subsection{Perdew-Zunger SIC}

As already mentioned in the introduction, a very popular approach to the 
definition of a one-particle self-interaction error and a corresponding 
correction was presented by Perdew and Zunger \cite{Per81}.  
The self-interaction error is given 
by accumulating the contributions from the individual orbital
densities $n_i({\bf r}) = |\varphi_i({\bf r})|^2$ for a set of
single-particle states $\varphi^N=(\varphi_1, \dots, \varphi_N)$. It
reads
\begin{subequations}
\begin{equation}
E_{\rm SI}[\varphi^N] = \sum_{i=1}^N \left(E_{\rm H}[n_i] + E_{\rm xc}[n_i]\right) \quad ,
\end{equation}
where $E_{\rm H}$ is the Coulomb Hartree energy and $E_{\rm xc}$ the
density functional for exchange and correlations.  Note that this is
not a functional of density alone. In fact, $E_\mathrm{SI}[\varphi^N]$
depends on the detailed orbitals.  
The PZSIC is defined by subtracting the self-interaction error from the
original functional, i.e.,
\begin{equation}
  E_{\rm PZSIC}[\varphi^N] = E_{\rm H}[n]+E_{\rm xc}[n] -  E_{\rm SI}[\varphi^N]
  \quad,
\label{eq:PZ-SIC}
\end{equation}
\end{subequations}
where $n=\sum_{i=1}^N n_i$ is the total electronic density.

The mean-field equations are derived in straightforward manner by
variation of the SIC energy $E_{\rm PZSIC}$ with respect to the
occupied single-particle orbitals $\varphi_i$.  It turns out that
mean-field Hamiltonian depends explicitly on the particular single-particle 
state on which it acts. This emerges because the PZSIC energy functional is not
invariant under unitary transformations amongst the occupied
states. There are several ways to deal with such a state-dependent
Hamiltonian \cite{Ped84,Goe97,Polo,Vyd04}. A particularly efficient
way is to use two different sets of single-particle states which are
connected by a unitary transformation amongst occupied states. That is
actually the solution scheme which we are using, for details see
\cite{Mes08c}. In this approach, the HOMO level is defined as usual,
in the basis-set which diagonalizes the Hamiltonian matrix.

\subsection{Average Density SIC}

The average density SIC (ADSIC) starts from the SIC energy (\ref{eq:PZ-SIC}) and simplifies it
by assuming that indistinguishable electrons are represented by equal 
single-particle densities.
In such an extreme simplification, one expresses them as the one-particle
fraction of the total spin-density
$n_i(\mathbf{r})=n_{\sigma_i}(\mathbf{r})/N_{\sigma_i}$ where
$\sigma_i$ is the spin of state $i$ and $N_{\sigma_i}$ the number of
particles with spin ${\sigma_i}$. In such a scheme, the standard PZSIC 
functional is represented by the ADSIC functional~:
\begin{eqnarray}
  E_{\rm ADSIC}[n_\uparrow,n_\downarrow] 
  &=& 
  E_{\rm H}[n]+E_{\rm xc}[n] 
\nonumber\\
  &&
  - 
  \sum_{\sigma\in\{\uparrow,\downarrow\}}{N_\sigma}
  \left(E_{\rm H}[n_\sigma]+E_{\rm xc}[n_\sigma]\right)
\label{eq:ADSIC}
\end{eqnarray}
where $n=n_\uparrow+n_\downarrow$. This is a spin-density functional
and can be treated in the same manner as any LDA or GGA scheme. 
This makes it extremely simple and efficient to use in atomic and molecular systems.  
However, the ADSIC functional contains a cumbersome
non-locality as it explicitly depends on the particle number
$N_\sigma=\int\!d^{3}\mathbf{r}\ n_\sigma({\bf r})$. This inhibits an
application in periodic systems, where $N_\sigma$ is infinite.

\section{Results}

\subsection{Numerical scheme and pseudo-potentials}
The calculations use a representation of the single-particle
wave functions on a coordinate-space grid with a spacing of 0.2 \AA. 
Densities and fields where represented on a refined grid of
0.1 \AA\ to account for the higher Fourier components in products
of single-particle states. The core electrons are handled within the 
frozen-core approximation by a real-space implementation of the projector
 augmented wave (PAW) method \cite{Blo94} using a development version of GPAW 
\cite{Mor05}.  The projectors and partial waves of the PAW method are 
taken as provided within the GPAW repositories for bare LDA exchange and 
correlation, i.e. without accounting for a SIC. 

This corresponds to use pseudo-potentials developed for LDA applications
in the context of PZSIC or ADSIC without readjustment of the pseudo-potential
parameters. This minor inconsistency is acceptable 
in various applications of SIC \cite{Par11, Alv11, Hof12}. 
Here such an improvement is avoided 
in favor of using a unique set of pseudo-potentials for all energy 
functionals. 

For the following survey, we show results from LDA using the PW92
parameterization \cite{Per92}. For most of the examples below, we have
also performed GGA calculations with the PW91 functional \cite{Per91}.
Even if the GGA slightly improves the overall quality of the IP, 
in particular if calculated from energy differences, with very few exceptions, 
the effect of the gradient dependence is less than 0.5 eV. Thus it neither 
affects the overall magnitude of errors or change the general trends that 
are discussed in the following sections. We therefore focus on the LDA part 
in this survey.

\subsection{Atoms}
The first step is to investigate the performance of both SIC approaches 
for atoms. The latter ones are the basic building blocks of molecules and solids.
Thus they must be correctly described before we can proceed to more
complex scenarios. The electronic structure of atoms incorporates
single-electron states with similar shape but different spatial extensions.
Thus atoms are a critical test case for SIC which is known to strongly depend
on the level of localization. 

\begin{figure}[htbp]
 \begin{center}
\includegraphics[width=0.95\columnwidth]{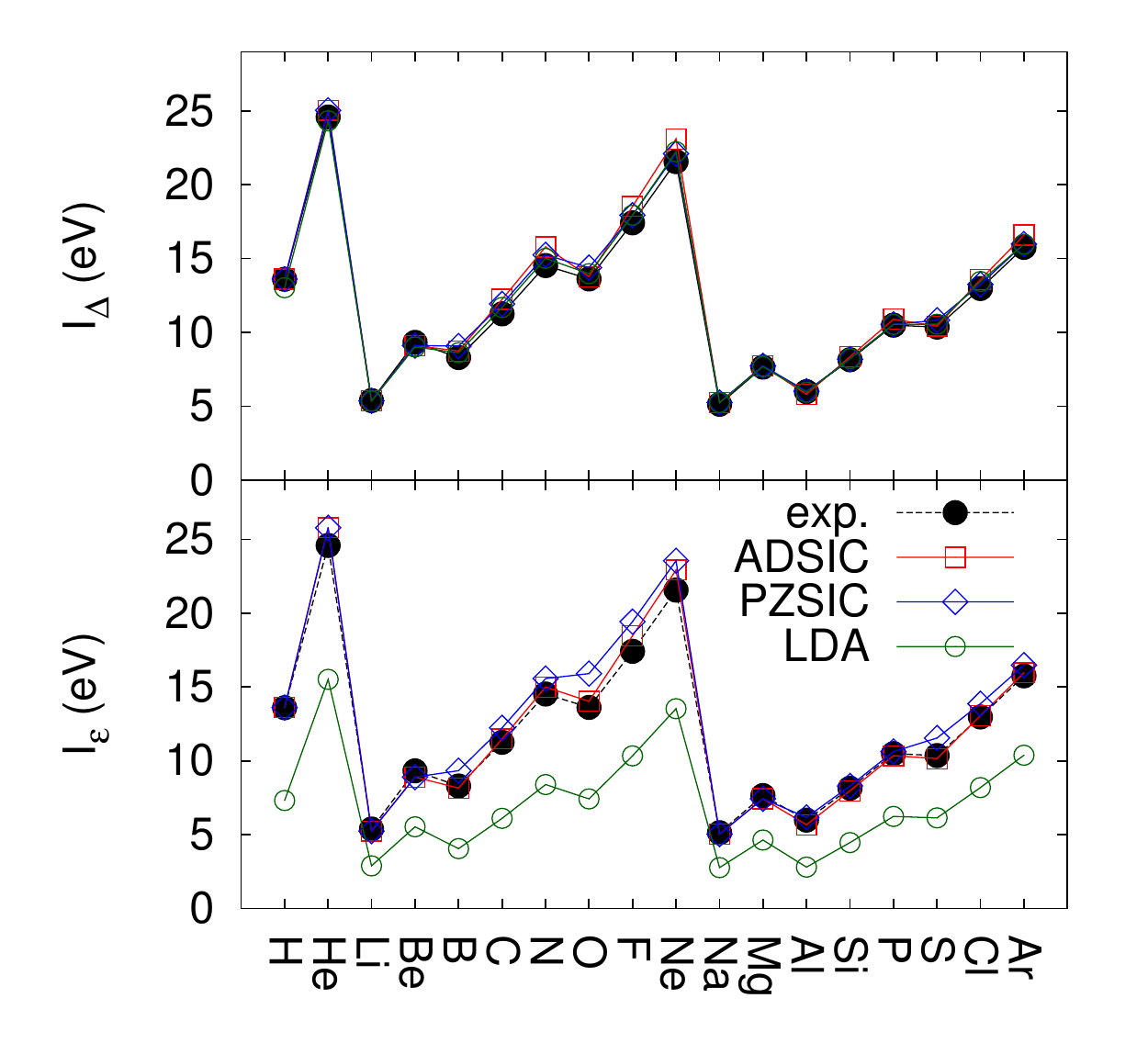}
 \end{center}
 \caption{\label{fig:IP_atoms}(Color online) Ionization potentials, $I_\Delta$ from 
   Eq.~(\ref{eqn:IP_ediff}), and $I_\varepsilon$ from Eq.~(\ref{eqn:IP_HOMO}),
   for neutral atoms from hydrogen to argon 
   and different approaches to self-interaction correction~:
   Average density SIC (squares), Perdew-Zunger SIC (diamonds), and
   the uncorrected local-density approximation (open circles). Experimental data
are displayed as closed circles \cite{cccbdb}.}
\end{figure}

 
Figure \ref{fig:IP_atoms} shows the IP as such for neutral atoms from
hydrogen ($Z=1$) to argon ($Z=18$). 
All methods yield very similar IP if 
it is evaluated as $I_\Delta$, i.e. as the energy difference
(\ref{eqn:IP_ediff}). Results differ more for
$I_\varepsilon$ computed from the HOMO according to
Eq.~(\ref{eqn:IP_HOMO}).  Here, the bare (semi-)local energy
functionals underestimate the ionization energy of by 30-40\%. The
defect is well known and can be traced back to the wrong asymptotic
behavior of the exchange-correlation potential for $|{\bf r}|\to\infty$ 
\cite{Per81}. 
Obviously, both SIC approaches cure the problem and
yield excellent agreement with experimental data.  
ADSIC is slightly superior in case of open shell atoms, while PZSIC 
slightly overestimates the IP.
Accounting for GGA (not shown here) has an insignificant effect for 
both $I_\Delta$ as well as $I_\varepsilon$. The results do not sufficiently 
differ to justify a separate plot.

\begin{figure}[htbp]
\begin{center}
\includegraphics[width=0.95\columnwidth]{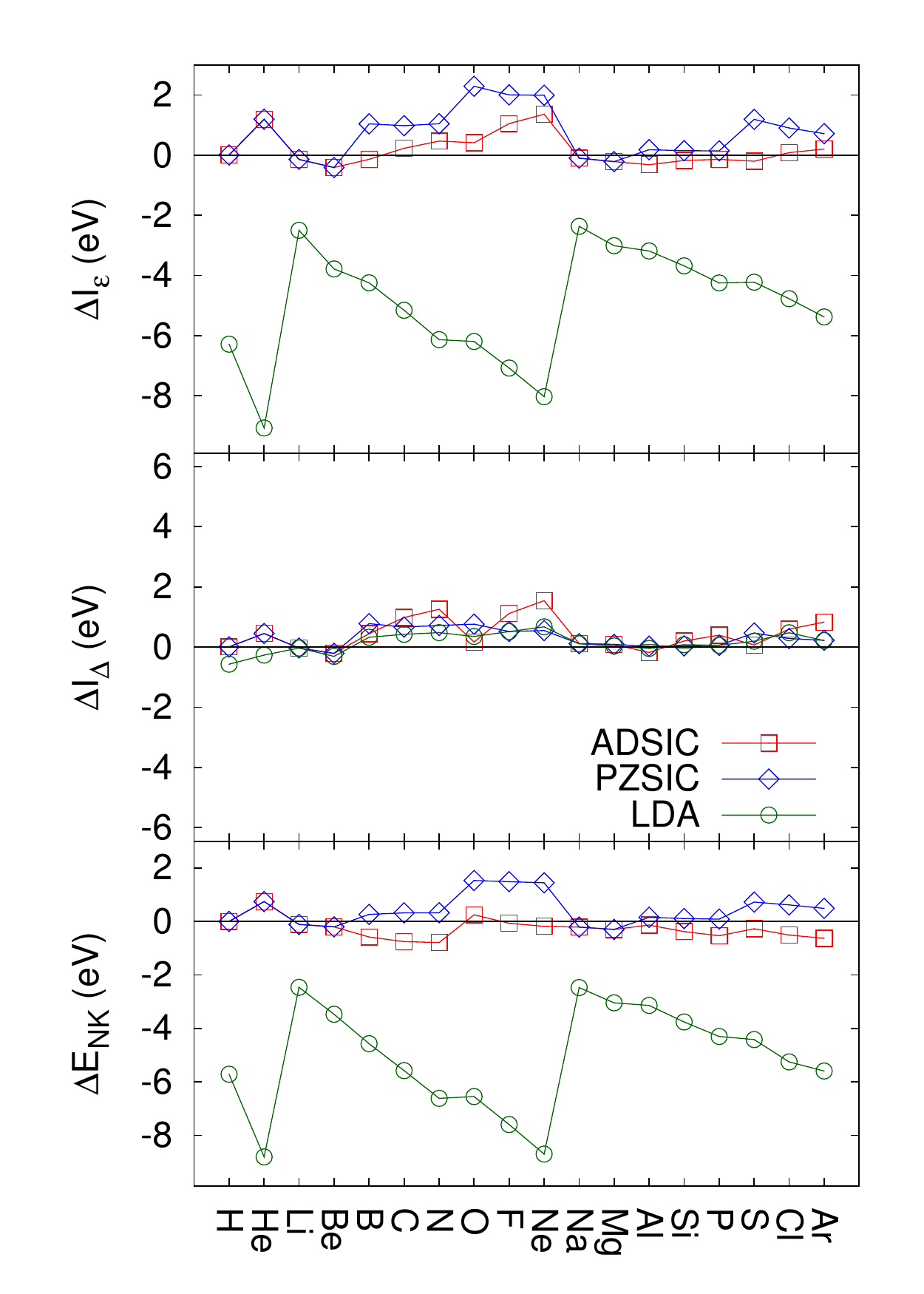}%
 \end{center}
\caption{\label{fig:IP_atoms_errors}(Color online) Errors in calculated IP compared 
with experimental values for the series of
  atoms depicted in Figure \ref{fig:IP_atoms}. Upper and middle panels~: 
  error from $I_\varepsilon$ and $I_\Delta$ respectively, according to Eq.~(\ref{eqn:errors}). Lower panel~:
  non-Koopman's energy defined in Eq.~(\ref{eqn:NK}).  }
\end{figure}

Figure \ref{fig:IP_atoms_errors} shows the same data of figure
\ref{fig:IP_atoms} but in terms of errors with respect to experimental 
data and of the NK energy. 
This reveals some differences between PZSIC and 
ADSIC where, somewhat surprisingly, the technically much simpler ADSIC 
visibly yields smaller NK energies and errors $\Delta I_\varepsilon$.

At this point, it is worth recalling that the ADSIC scheme can be derived
 as an approximation to PZSIC, assuming a most delocalized representation
of the single-particle densities. In ADSIC, orbitals are extending 
over the whole atom in stark contrast to the localized orbitals 
that are commonly found in PZSIC calculations \cite{Klu11}. It thus appears 
that a significant higher level of delocalization is actually desirable, 
which confirms previous concerns that PZSIC orbitals are too localized.

\subsection{Simple Molecules}
As a next step, we consider simple molecules, as many dimers, and a few
more complex ones. The selection has been adapted from \cite{Per96}.
It covers systems which do not have the problem of spatial symmetry 
breaking by an unrestricted mean-field calculation. Reference data was 
taken from \cite{cccbdb}.

\begin{figure}[htbp]
 \begin{center}
 \includegraphics[width=0.95\columnwidth]{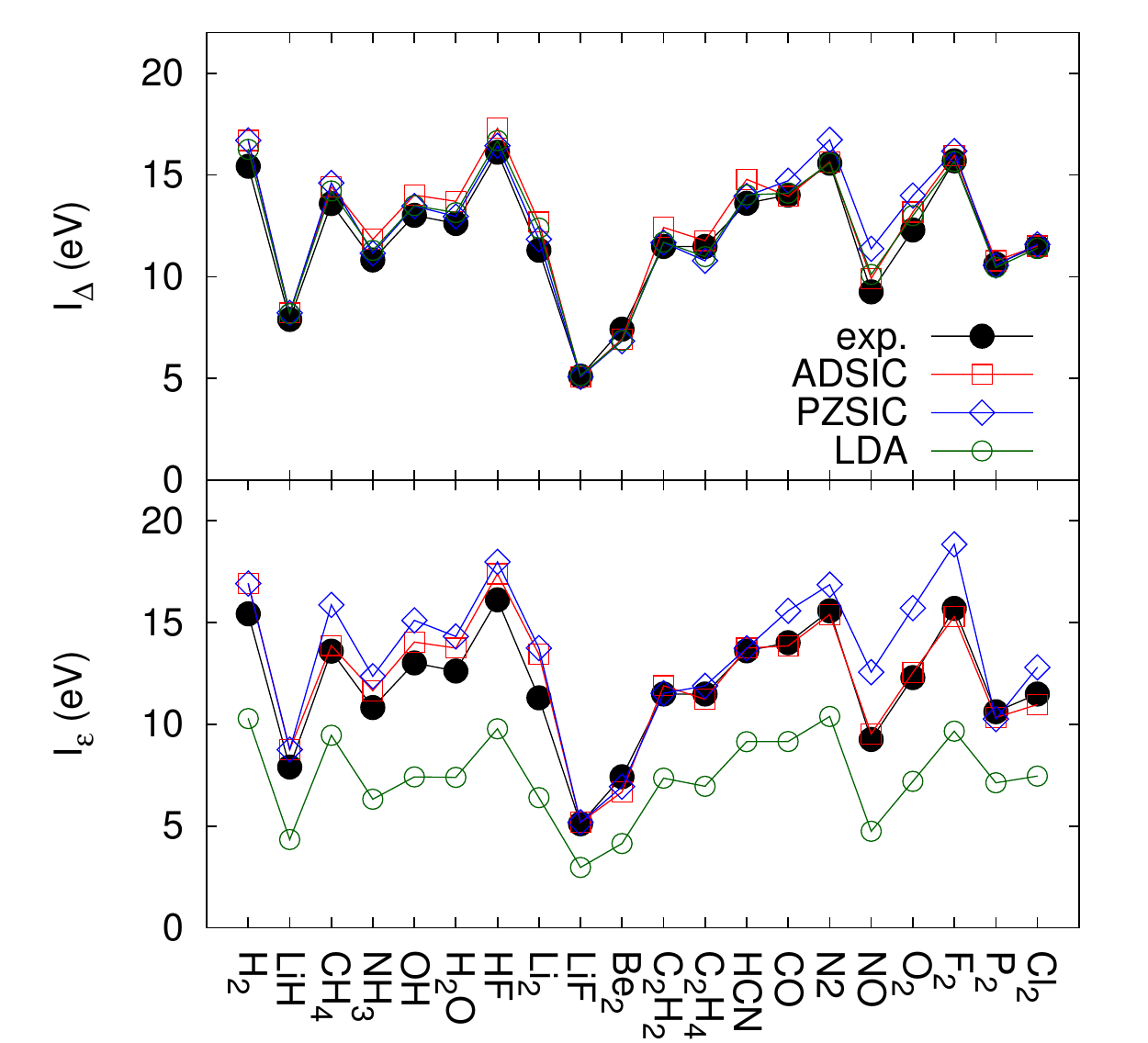}%
 \end{center}
 \caption{\label{fig:IP_molecules}(Color online) Same as in Figure \ref{fig:IP_atoms},
   but for a set of simple molecular systems.}
\end{figure}

Figure \ref{fig:IP_molecules} shows the IP for a chosen set of
molecules.
At first glance, the results resemble those for atoms in
Figure \ref{fig:IP_atoms}. The $I_\Delta$ provides reasonable results
for all methods while $I_\varepsilon$ shows dramatic differences
between LDA and the SIC models. However, taking a closer look, we also see that
results for ADSIC and PZSIC show larger differences than in the case of
atoms. Somewhat surprisingly, ADSIC comes again much closer to experimental
data than PZSIC.

\begin{figure}[htbp]
 \begin{center}
 \includegraphics[width=0.95\columnwidth]{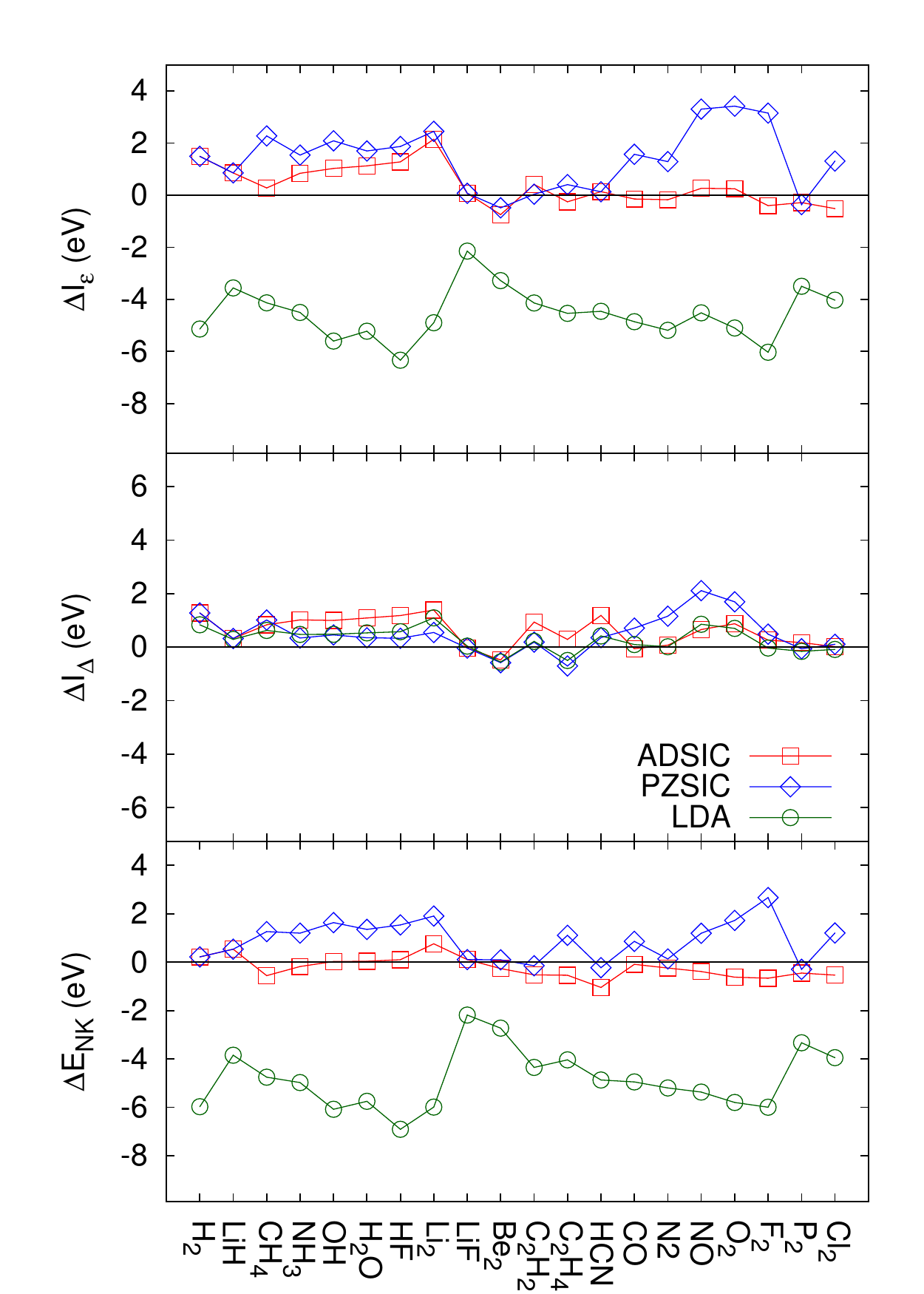}
 \end{center}
 \caption{\label{fig:IP_molecules_errors}(Color online)
   Same as in Figure \ref{fig:IP_atoms_errors}, but 
   for a set of simple molecular systems.}
\end{figure}

Figure \ref{fig:IP_molecules_errors} shows the data from the previous
figure in terms of energy differences, the errors $\Delta I$ as
compared to reference data and the NK energy.
The results confirm the
impressions indicated in the comparison of the IP as such~: The
$\Delta I_\varepsilon$ shows significant differences between PZSIC and
ADSIC since the latter one generally performs better.

One may argue that comparison with reference data is also influenced 
by other details of the calculations or the choice of the reference
data. The NK energy (lowest panel) is free of these uncertainties. ADSIC
clearly delivers the smallest NK energies. This was seen already for
atoms. But here in the case of molecules the effect is even more
pronounced as PZSIC shows larger deviations.

\subsection{Systematic sets of molecules}
In this section, we look at a systematic variation of molecules around
basic carbohydrates.  The first family ($\textrm{CH}_x$) represents a
variation of the number of C-H bonds. The second family changes the
character (single, double, triple bonds) in
$\textrm{C}_2\textrm{H}_n$.  The third family is similar to the first
one but replacing carbon by a heavier element (silicon) with the same
number of valence electrons, while the fourth series replaces the
carbon atom by nitrogen (which has a different valence). In the final
series one of the single bonded hydrogen atoms in $\textrm{CH}_4$ is
substituted by a different group.  The last element of the third and
fourth series are not strictly within the systematics.

%
%

\begin{figure}[htbp]
 \begin{center}
 \includegraphics[width=0.95\columnwidth]{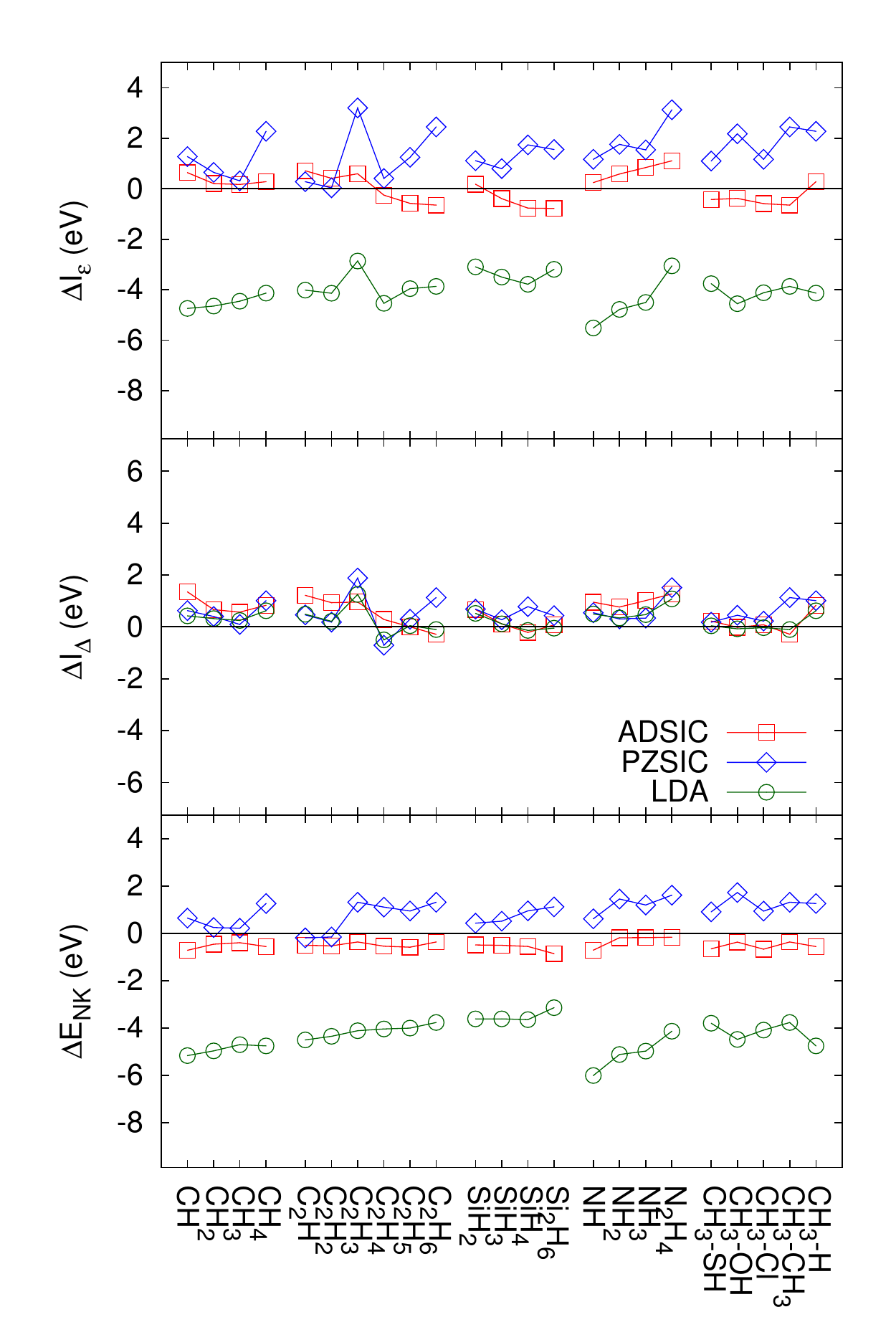}%
 \end{center}
 \caption{\label{fig:IP_chainmol_errors}(Color online) Same as in Figure
   \ref{fig:IP_atoms_errors}, but for the families of molecules with
   systematically varied properties.}
\end{figure}

We have seen in the previous systems that energy differences are
showing more details than the energies as such. We thus proceed here
immediately to energy differences which are compiled in Figure
\ref{fig:IP_chainmol_errors}.
The results are very similar to the
previous case of simple molecules, see Figure
\ref{fig:IP_molecules_errors}. Some of the deviations are however larger than
in the previous case. This indicates that these complex molecules are
more critical test cases. Even in this more demanding scenario, we
find again that ADSIC performs superior with respect to the
deviation from reference data and even more so for the NK energies.

Thus we find that the ADSIC which assumes orbital densities
that are delocalized over the whole molecule yields a systematic 
improvement over PZSIC. This is somehow surprising in view of the deficits of 
ADSIC, in particular as its inability to describe dissociation and the lack of size 
consistence are directly attributed to a too high level of delocalization. 

For infinite matter, ADSIC is not applicable due to the explicit dependence on the particle number. 
Already for larger systems, the explicit dependence on the total particle 
number quickly renders the SIC contribution to the energy functional an 
inefficient approach to cure problems of the LDA.
The observation that delocalization on the length-scale of small molecules is 
in fact favorable for the quality of the NK energy and IP calls for more 
systematic investigations.

\subsection{Carbon rings and chains}
The self-interaction error on the IP for the Coulomb Hartree term is
typically of order of $e^2/R$ where $R$ is the radius of the system.
The error for the exchange-correlation potential can be estimated 
within ADSIC as $v_\mathrm{xc}[n/N]$. Both shrink with increasing
system size.
%
%
In order to explore the evolution of the self-interaction errors with
increasing size, we consider carbon rings and chains. For the latter ones,
we only consider odd numbers of atoms because only these have stable
electronic configurations for spin saturated ground states. The
carbon atoms have more or less constant bond length.
This means that increasing the number of carbon atoms induces a (linear) 
growth of the geometrical extension, either of the chain or the ring.

\begin{figure}[htbp]
 \begin{center}
\includegraphics[width=1.0\columnwidth]{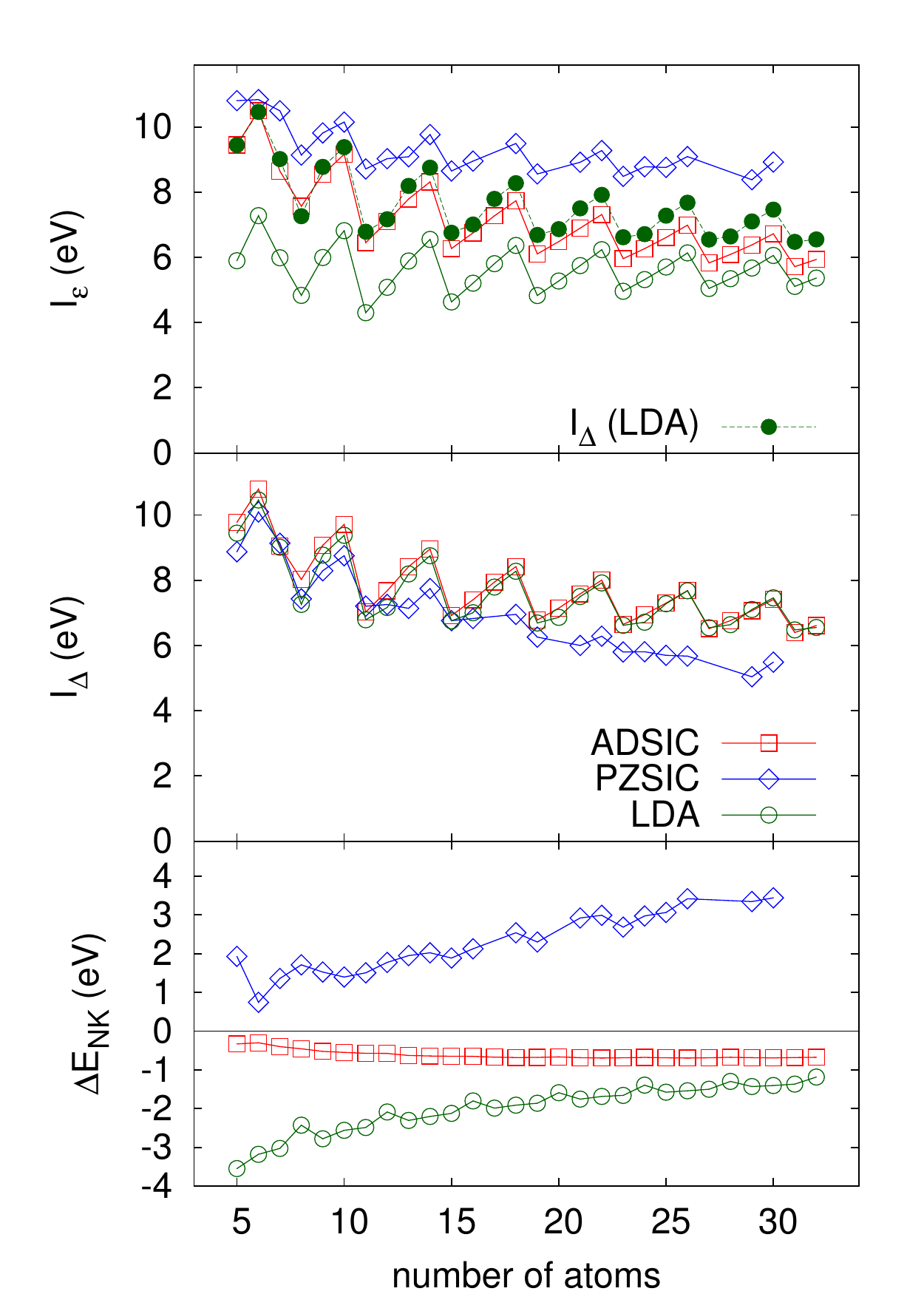}%
  \end{center}
 \caption{\label{fig:carbon_rings}(Color online) Non-Koopmans energies (bottom), and 
ionization potentials $I_\Delta$ (from energy differences, middle) and 
$I_\varepsilon$ (from the HOMO, top), computed in
   various schemes for carbon rings of various size ($5\leq
   N_\textrm{atoms}\leq 32$). In the top panel, $I_\Delta$ from LDA (see 
middle panel) is 
superimposed to the $I_\varepsilon$ calculated in LDA, ADSIC and PZSIC.}
\end{figure}

The upper two panels of Figure \ref{fig:carbon_rings} show the IP for
carbon rings as a function of the number of atoms.
Comparing ADSIC and LDA, we see again the equally good performance 
for $I_\Delta$, and the large
self-interaction error in $I_\varepsilon$ for LDA while ADSIC behaves
very well. The reference data, here calculated LDA values $I_\Delta$, 
show a pronounced step structure due to the successive filling of the
electronic shells. Large $I$ indicate particularly stable electronic 
structures, i.e., shell closures. The sudden
reductions shows that a new, and less bound, electronic shell has to
be opened to place the given number of electrons. LDA and ADSIC
reproduce the shell effect.  On first glance, the PZSIC results are
quite surprising since they deviate even qualitatively from the other
results, as they show less pronounced shell effects, at least
with increasing chain length. 

It shall be noted, that missing points also indicate that reliable
minimization of the PZSIC energy becomes challenging for anionic 
configurations, where various local minima exist. The local minima correspond
to different, almost energetically equivalent, configurations with 
different levels of delocalization of the excess electron in the 
spin-majority channel. 
The effect is worse for mid-shell systems but less problematic for closed shell
configurations. No such problem exists for ADSIC due to the absence of the
orbital-dependence.

The lowest panel of Figure \ref{fig:carbon_rings} shows the NK
energies. The $\Delta E_\mathrm{NK}$ from LDA starts large but shrinks
with increasing size as one could have expected. The ADSIC result is
small throughout, but has a slight tendency to increase with size, and of
course, never becoming larger than the error from LDA. However, the
$\Delta E_\mathrm{NK}$ from PZSIC is generally large and even grows with
system size. This finding is rather cumbersome, as it confirms that 
the difference between the behavior of LDA and ADSIC on the one hand,
and PZSIC on the other hand, actually stems from misconceptions in the PZSIC
partially compensated in the approximate ADSIC.

The significant and positive NK energy indicates that strong
correlation effects, which are underestimated in semi-local exchange and
correlation, are overestimated by the PZSIC. The screening of such strong 
correlation effects has to be reintroduced in the self-interaction corrected
approach, e.g., by the assumption of more delocalized states, as in case of
the ADSIC.

The convergence of ADSIC and LDA yet illustrates the collapse of ADSIC 
as a working SIC scheme for extended systems, where $N\gg1$, contrary to the
case of small $N$ where the NK energy is still improved significantly. 
The almost constant but finite NK energy indicates that, although ADSIC is 
not capable of a complete curing of the non-linear dependence of the 
LDA energy functional for fractional occupation, it at least provides a 
scheme that yields similar magnitudes of errors for compact systems and 
extended ones, whenever LDA is by itself considered a reasonable 
approximation there.

\begin{figure}[htbp]
 \begin{center}
 \includegraphics[width=1.0\columnwidth]{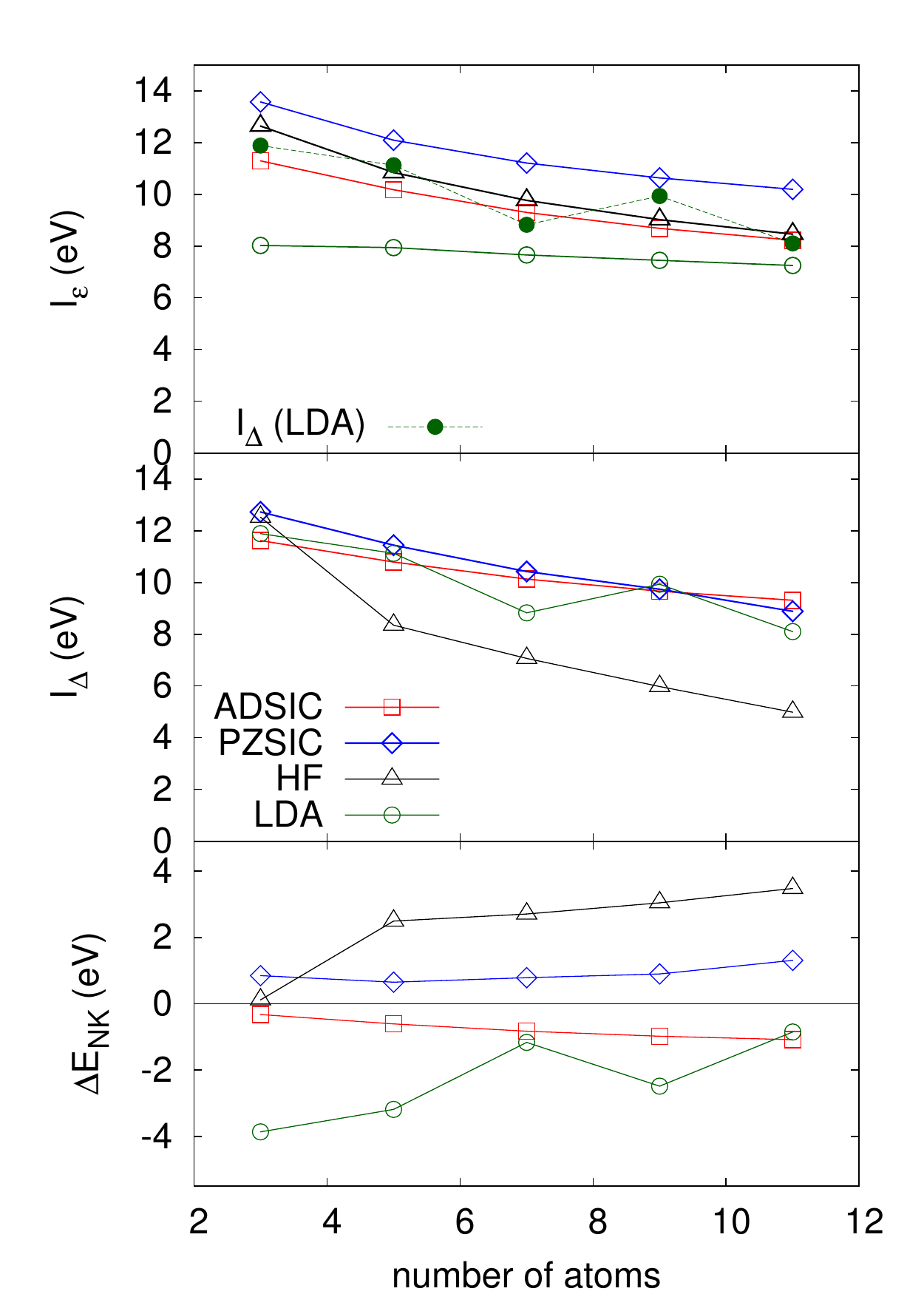}%
 \end{center}
 \caption{\label{fig:formation_carbon}(Color online) Same as in Figure \ref{fig:carbon_rings} but
for linear carbon chains ($3\leq N_\textrm{atoms}\leq 11$). The data is complemented 
by results obtained from bare exchange-only Hartree-Fock calculations (triangles).
}
\end{figure}

Figure \ref{fig:formation_carbon} shows IP and $\Delta E_\mathrm{NK}$
for carbon chains.
In this case, we added also results from a pure
Hartree-Fock (HF) calculation. PZSIC
looks more agreeable here than for rings. But note that we have considered
rather short chains. There are again large differences
between PZSIC and ADSIC. This time, however, they are distributed
almost symmetrically around zero error (see lowest panel). No clear
preference can be deduced in this example. 

The largest errors appear here for exact exchange in HF.  Starting out
perfect for the smallest chain $C_3$, The NK energy jumps already for
$C_5$ and continues to grow further. This sounds, at first glance,
very surprising as HF is free of any self-interaction error. However,
removal of one electron causes polarization effects on the mean field
of the remaining electrons. These may be small in compact molecules.
But polarizability grows huge particularly in chains. The missing
correlations from polarization effects are the source of the
increasing NK error with HF. This can also be seen from the IP as
such. The HF result deviates much from the others for $I_\Delta$
(middle panel), while it nicely stays  in between for $I_\varepsilon$.
The missing polarization effects explain the mismatch for HF.  The
fact that the DFT based methods perform better indicate that some
polarization effects are properly modeled in DFT, although it is also
known that DFT underestimated the polarizability in some chain
molecules \cite{Gis98}.

\subsection{Discussion}

To summarize the results presented in the above figures, we have
computed average errors for each group of system considered, atoms,
simple molecules, and families of systematically varied molecules.
Thereby, we distinguish between mean error, mean absolute error
and the error fluctuations defined as
\begin{subequations}
\label{eq:ME}
\begin{eqnarray}
  \mathrm{ME}(\Delta\mathcal{O})
  &=&
  \frac{1}{N_{\rm samp}}\sum_i \Delta\mathcal{O}_i
\quad, 
\\
  \mathrm{MAE}(\Delta\mathcal{O})
  &=&
  \frac{1}{N_{\rm samp}}\sum_i\left|\Delta\mathcal{O}_i\right| \quad,
\\
  \sigma(\Delta\mathcal{O})
  &=&
  \frac{1}{N_{\rm samp}}\sum_i\left|\Delta\mathcal{O}_i -
             \mathrm{ME}(\Delta\mathcal{O})\right| \quad,
\end{eqnarray}
\end{subequations}
where $\mathcal{O}$ is one of the considered observables, that is $I_\Delta$,
$I_\varepsilon$, or $E_\mathrm{NK}$. The index $i$ runs over the
$N_{\rm samp}$ samples in a given group, and $\Delta\mathcal{O}_i = \mathcal{O}_i-
\mathcal{O}_i^\mathrm{(ref)}$ stands for the observables deviation from the reference data 
$\mathcal{O}_i^\mathrm{(ref)}$. 
The resulting averages for each group are listed
in Table \ref{tab:IP}. 
\begin{table}[htbp]
 \begin{center}
  \begin{tabular}{lr|rrr|rrr|rrr}\hline\hline
&                   &\multicolumn{3}{c|}{$I_\Delta$} &
                    \multicolumn{3}{c|}{$I_\epsilon$} &
                    \multicolumn{3}{c}{$E_{\rm NK}$} \\
&                   &     ME   &  MAE &  $\sigma$    &   ME &  MAE &  $\sigma$    &   ME &  MAE &  $\sigma$ \\\hline
&atoms &&&&&&&&&\\                                                                         
&  LDA              &     0.2  & 0.3  &  0.3 & -5.0 & 5.0  &  1.5 & -5.1 &  5.1 &  1.6 \\
&  PZSIC           &     0.3  & 0.3  &  0.3 &  0.7 & 0.8  &  0.7 &  0.4 &  0.5 &  0.5 \\
&  ADSIC           &     0.4  & 0.5  &  0.4 &  0.2 & 0.4  &  0.4 & -0.3 &  0.4 &  0.3 \\\hline
&small molecules &&&&&&&&&\\                                                                       
&  LDA              &     0.3  &  0.4 &  0.4 & -4.6 &  4.6 &  0.7 & -4.9 &  4.9 &  1.0 \\
&  PZSIC           &     0.5  &  0.6 &  0.5 &  1.4 &  1.5 &  0.9 &  0.9 &  1.0 &  0.7 \\
&  ADSIC           &     0.6  &  0.7 &  0.5 &  0.4 &  0.6 &  0.6 & -0.2 &  0.4 &  0.3 \\\hline
&systematic mol. &&&&&&&&&\\                                                                  
&  LDA              &     0.3  &  0.4 &  0.3 & -4.1 &  4.1 &  0.5 & -4.3 &  4.3 &  0.5 \\
&  PZSIC           &     0.5  &  0.6 &  0.4 &  1.4 &  1.4 &  0.7 &  0.9 &  0.9 &  0.4 \\
&  ADSIC           &     0.5  &  0.6 &  0.4 &  0.1 &  0.5 &  0.5 & -0.5 &  0.5 &  0.1 \\\hline\hline
  \end{tabular}
 \end{center}
  \caption{\label{tab:IP}Mean error (ME), mean absolute error (MAE) and
    error fluctuations $\sigma$ as defined in eqs. (\ref{eq:ME}) for 
    IP as well as NK energy for the data sets shown in figures 
    \ref{fig:IP_atoms_errors},
    \ref{fig:IP_molecules_errors} and
    \ref{fig:IP_chainmol_errors}. Redundant data in
    \ref{fig:IP_chainmol_errors} is only considered once in the
    averages.}.
\end{table}
Computation of IP as $I_\Delta$,
i.e. from energy differences, is always a safe procedure yielding
reliable results already with LDA. 
%
Computation as $I_\varepsilon$ via the HOMO is possible with good accuracy 
in both SIC models. The great surprise is that the very simplistic ADSIC 
approach performs very well for the $I_\varepsilon$, typically even 
better than PZSIC. The same conclusion is deduced from the
non-Koopmans energy $\Delta E_\mathrm{NK}$. This was already seen from
the above figures and is corroborated in Table \ref{tab:IP} on a
quantitative level.

The excellent performance of ADSIC from compact systems both in terms
of accuracy and the small violation of Koopmans' condition is
remarkable. Still, one should keep in mind the known deficiencies of the
approach. Most notably is the violation of size consistency which
becomes apparent in the dissociation of a molecule. Consider a dimer
with total electron number $N$ which dissociates into one part containing
$N_1$ electrons and another one with $N_2$ electrons.  The ADSIC for the compound
involves, of course, the total electron number $N$. Since we follow the
dissociation path continuously, we necessarily have to keep using
$N$ in the correction. After all, we end up with two isolated atoms
which would be treated by one common correction still regulated by the
total $N$. This is, of course, wrong as we know that each single atom
has to be separately corrected with its own $N_i$.  The case is even
worse in violent dynamics leading to multi-fragmentation. The problem
could already have been spotted from the fact that the dependence on
$N=\int d^3r\,n(\mathbf{r})$ implies a non-locality which becomes
increasingly itching if $n(\mathbf{r})$ ceases to be compact, but is
rather distributed over several regions of space.

Fully accomplished dissociation and multi-fragmentation are, of
course, extreme limits. The defects of ADSIC in this respect tend to
show up earlier, for example, in the Born-Oppenheimer energies along
the dissociation path. Thus one should not use ADSIC for computing
large-amplitude molecular vibrations without careful checking its
range of validity for the given application. Problems may also show up
in molecules which combine very different length scales as, e.g., in
NaH$_2$O where the Na atom adds a rather dilute electron distribution
to the otherwise compact H$_2$O. In spite of the encouraging results
presented above, one should check the NK energy $\Delta E_\mathrm{NK}$
for each new application again. 

These known shortcomings should not hinder us to appreciate
the good performance attained by ADSIC in structural and low energy
dynamical situations. As illustrated all along the present work, ADSIC provides a 
remarkably robustness in terms of Koopmans'
violation. This implies, in particular, that it can be safely used in the 
perturbative dynamical regime where only a tiny fraction of an electron is emitted.

\section{Summary and Conclusions}
We have compared the performance of two different approaches to self-interaction
correction regarding calculated ionization potentials and violation of Koopmans' theorem.
We have focused the discussions on two SIC procedures: the original Perdew-Zunger
approach (PZSIC) and the average density version thereof (ADSIC).
A wide range of electronic systems has been considered ranging from atoms and simple molecules 
up to systematics of moderate size molecules, in particular carbon systems. The overall survey is 
thus quite general, so that the conclusions attained have a safe ground, beyond any specific effect.

We find in all examples considered here that ADSIC provides more reliable estimates of IP and
a smaller violation of Koopmans' theorem. This is a welcome result in view of the 
remarkable simplicity (and correlatively low computational price) of ADSIC.

We have also explored a known collapse of the ADSIC approach for extended
systems. It was shown that PZSIC also fails to cure
flaws of LDA in this regime. 
An optimistic interpretation of the data obtained on the
example of carbon chains allows to conclude that an efficient orbital-density dependent SIC 
should provide weak localization of the single-electron states over
several atoms. Such a weak localization is in line with the excellent performance of bare ADSIC
in case of the smaller molecules
studied here.
%

Whereas the results of this survey question the quality of PZSIC as a benchmark approach
to a SIC, they simultaneously encourage the educated use of the much
simpler ADSIC approach. 
However, as also noted, ADSIC certainly does not provide
the ultimate SIC scheme as it fails by construction, for example in the modeling of
dissociation processes or strong ionization. The limits of ADSIC with respect to
dissociation or molecular structural rearrangement need to be 
explored further. 
Still it remains a viable and robust option for many dynamical situations, especially in the case
of perturbative ionization, where the ionization potential, precisely the negative HOMO
level, plays a central role. This implies that ADSIC remains the favorably
self-interaction correction in the calculation of reliable photo-electron spectra and
angular distributions of emitted electrons, which represent an ever-growing issue in the
dynamics of irradiated clusters and molecules.

Future work should also aim at investigating to which extent the level
of localization can be controlled within the PZSIC scheme by modifying the
functional form, e.g., within the framework of GGA-SIC or by implying 
alternative localization criteria during the optimization of internal degrees 
of freedom, i.e., the unitary transformation amongst the 
single-particle states. 

\begin{acknowledgments}
The authors acknowledge support from Institut Universitaire de France. One of us (PK)
also thanks the Laboratoire de Physique Theorique de Toulouse
for its hospitality, the Centre National de la Recherche Scientifique for
financial support, and P. Wopperer and S. Kl\"upfel for fruitful 
discussions. Allocations of computational resources at the Regional Compute 
Center Erlangen (RRZE), Calcul en Midi-Pyr\'en\'ees (CALMIP) and under the 
Nordic High Performance Computing (NHPC) project are gratefully acknowledged.
\end{acknowledgments}

\bibliography{koopmans}

\end{document}